\documentclass[pra,twocolumn]{revtex4}
\usepackage{epsfig}

\newtheorem{definition}{Definition}
\newtheorem{lemma}{Lemma}
\newtheorem{theorem}{Theorem}

\begin{document}

\title{Reduction Theorems for Optimal Unambiguous State Discrimination of
Density Matrices}

\author{Philippe Raynal}

\author{Norbert L\"utkenhaus}

\affiliation{Quantum Information Theory Group, ZEMO, University
Erlangen-N\"urnberg, Staudtstr. 7/B2, D-91058 Erlangen, Germany}

\author{Steven J. van Enk}

\affiliation{Bell Labs, Lucent Technologies, 600-700
Mountain Ave., Murray Hill NJ 07974}

\date{\today}

\begin{abstract}

We present reduction theorems for  the problem of optimal unambiguous state discrimination (Optimal USD) of two general density matrices. We show that this problem can be reduced to that of two density matrices that have the same rank $n$ and are described in a Hilbert space of dimensions $2n$. We also show how to use the reduction theorems to discriminate unambiguously between $N$ mixed states ($N \ge 2$).

\end{abstract}

\maketitle

\section{Introduction}

Discrimination between quantum mechanical states is a standard task in quantum communication protocols. Typically, the signals are represented by non-orthogonal quantum states, often even in the form of mixed states due to the effect of noisy transmission channels. It is known that non-orthogonal pure states cannot be discriminated exactly, that means, with full efficiency without any error. However, it is possible to discriminate between them unambiguously in a certain fraction of the cases. This problem was formulated by Dieks \cite{dieks} and Ivanovic \cite{ivanovic} for two pure states with equal {\it a priori} probabilities, and was elegantly solved by Peres \cite{peres}. The more general scenario of two pure states with arbitrary {\it a priori} probabilities was investigated and solved by Jaeger and Shimony \cite{jaeger}. More general scenarios with a higher number of signal states were treated, with analytic results for signal states that satisfy certain symmetries \cite{terno}, \cite{chefles}. 

It is only recently that the problem of generalization to mixed states has been considered. One of the first works describes the optimal unambiguous discrimination between a pure state and a density matrix of rank $2$ \cite{hillery}, which has been generalized for a rank $n$ density matrix \cite{bergou}. In a more recent paper, Rudoph {\it et al.} \cite{rudolph} solved other special cases, for example that of two density matrices of rank $(n-1)$ in a $n$-dimensional subspace. Finally, necessary and sufficient conditions for optimality and some numerical methods are discussed by Fiurasek and Jezek \cite{fiurasek} and by Eldar \cite{eldar}.

In this article, we present two theorems that allow to reduce the general problem of optimal unambiguous state discrimination of two density matrices to standard forms. This standard problem can be formulated as optimal unambiguous state discrimination of two density matrices of rank $n$ that are defined in a minimal Hilbert space of dimension $2n$. We show that the previously given examples involving density matrices can be reduced effectively to the optimal unambiguous state discrimination of two pure states, as solved by Jaeger and Shimony \cite{jaeger}. The next challenge is therefore to discriminate two density matrices with a two dimensional rank in a four dimensional Hilbert space.\\

The paper is organized as follows. In section II, we present the mathematical background of the unambiguous state discrimination and introduce some notations. Then, in section III and IV, based on a geometrical approach, we present two theorems to simplify the problem of unambiguous state discrimination between two arbitrary density matrices. In section V, we propose a brief discussion on the consequences of the two previous theorems including a generalization to discrimination between more than two density matrices. Finally, we present our conclusions in section VI.

\section{Optimal Unambiguous State Discrimination}

If quantum states cannot be discriminated exactly, one can search for an optimal distinction between them. The meaning of ``optimal'' needs to be made precise, and in quantum state estimation theory \cite{helstrom} this is typically expressed via cost-functions. These cost functions provide a figure of merit for different measurements. Optimal unambiguous state discrimination is an extreme case in that we are looking for a measurement that either identifies a state uniquely (conclusive result) or fails to identify it (inconclusive result). The goal is to minimize the fraction of the inconclusive results. For the problem to be properly stated, one needs to fix the set of quantum states $\{\rho_i\}$ together with the corresponding {\it a priori} probabilities $\{p_i\}$ of their appearance. The measurements involved are typically generalized measurements \cite{kraus} which are described by a Positive Operator Valued Measure (POVM). A POVM is a set of positive semi-definite and hermitian operators $\{F_k\}$ that satisfies the completeness relation $\sum_k F_k = \openone$ on the Hilbert space spanned by the signals. The probability to obtain outcome $k$ for a given signal $\rho_i$ is then given by $p(k|i)={\rm Tr}(\rho_i F_k)$.
\\

\begin{definition}
A measurement described by a POVM $\{F_k\}$ is called an Unambiguous State Discrimination Measurement (USDM) on a set of states $\{\rho_i\}$ iff the following conditions are satisfied:
\begin{itemize}
\item The POVM contains the elements $\{F_?,F_1, \dots F_{N}\}$ where $N$ is the number of different signals in the signal set. The element $F_?$ is connected to an inconclusive result, while the other elements $F_i$ correspond to an identification of signal state $\rho_i$. 
\item No signals are wrongly identified, that is ${\rm Tr}(\rho_i F_k) = 0 \quad\quad \forall i \neq k \quad i,k=1,...,N$.
\end{itemize}
\end{definition}

Each USD Measurement gives rise to a failure probability, that is, the rate of inconclusive results. This can be calculated as

\begin{eqnarray}
Q[ \{F_k\}] := \sum_i p_i {\rm Tr}(\rho_i F_?) \; .
\end{eqnarray}

\begin{definition}
A measurement described by a POVM $\{F^{opt}_k\}$ is called an Optimal Unambiguous State Discrimination Measurement (OptUSDM) on a set of states $\{\rho_i\}$ with the corresponding {\it a priori} probabilities $\{p_i\}$ iff the following conditions are satisfied
\begin{itemize}
\item The POVM $\{F^{opt}_k\}$ is a USD measurement on $\{\rho_i\}$
\item The probability of inconclusive results is minimal, that is $Q[\{F^{opt}_k\}] = \min_{F\in USDM} Q[\{F_k\}]$.
\end{itemize}
\end{definition}

In this article, we are searching for an optimal USD measurement to discriminate  two arbitrary density matrices $\rho_1$ and $\rho_2$ that are prepared with {\it a priori} probability $p_1$ and $p_2$ respectively. We find that this general problem can be reduced to a simpler standard situation. Moreover, the reduction can be applied as well to the case of more than two density matrices, as seen later.\\

Firstly, let us fix some notations. The reduction theorems make use of the support ${\cal S}_{P}:={\rm support}(P)$ of positive semi-definite and hermitian operator $P$. The support of a positive semi-definite and hermitian operator is defined as the subspace spanned by eigenvectors of $P$ corresponding to non-zero eigenvalues. We denote also $r_P:={\rm rank}(P)={\rm dim}({\cal S}_{P})$, the rank of $P$. Next we define, in a Hilbert space $\cal H$, the sum and the intersection of two Hilbert subspaces ${\cal H}_1$ and ${\cal H}_2$. The sum ${\cal H}_1 + {\cal H}_2$ of the subspaces ${\cal H}_1$ and ${\cal H}_2$ is defined to be the set consisting of all sums of the form $a_1+a_2$, where $a_1 \in {\cal H}_1$ and $a_2 \in {\cal H}_2$. ${\cal H}_1 + {\cal H}_2$ is a Hilbert subspace of $\cal H$. The intersection ${\cal H}_1 \cap {\cal H}_2$ is defined to be the set consisting of all the elements $a$, where $a \in {\cal H}_1$ and $a \in {\cal H}_2$. ${\cal H}_1 \cap {\cal H}_2$ is a Hilbert subspace of $\cal H$.
The complementary orthogonal subspace (orthogonal complement) of a subspace $\cal S$ in $\cal H$, written ${\cal S}^{\perp}$, is the set of all the elements of $\cal H$ orthogonal to $\cal S$. We then have $\cal H=\cal S \oplus {\cal S}^{\perp}$, the direct sum of the two orthogonal subspaces. Finally, we denote by $\Pi_{\cal S}$ the orthogonal projection onto the subspace $\cal S$.

\section{Common subspace of the supports}

In the first theorem, we will consider the situation where the supports of the two density matrices have a common subspace. This is the case whenever we find that
\begin{eqnarray}
{\rm dim}\left({\cal S}_{\rho_1}\right) + {\rm dim}\left({\cal S}_{\rho_2}\right) >
{\rm dim}\left({\cal H}\right) \;
\end{eqnarray}
Here ${\cal H}$ is the Hilbert space spanned by the two supports. In this case, it can be written as
\begin{eqnarray}
{\cal H} = {\cal H'}\oplus {\cal H_\cap}
\end{eqnarray}
where ${\cal H_\cap}= {\cal S}_{\rho_1} \cap {\cal S}_{\rho_2}$ is the common subspace of the two supports, and ${\cal H'}$, its  orthogonal complement  in ${\cal H}$. The first reduction theorem will eliminate the common subspace ${\cal H_\cap}$ from the problem. The intuitive reason is that in this subspace no unambiguous discrimination is possible, so the population of the two density matrices on it will contribute always only to the failure probability, never to the conclusive results. This is made precise in the following theorem. \\


\begin{theorem}
Reduction Theorem for a Common Subspace\\

Suppose we are given two density matrices $\rho_1$ and $\rho_2$ on ${\cal H}$ with {\it a priori} probabilities $p_1$ and $p_2$ such that their respective supports ${\cal S}_{\rho_1}$ and ${\cal S}_{\rho_2}$ have a non-empty common subspace ${\cal H_\cap}$. We denote by ${\cal H'}$ the orthogonal complement of ${\cal H_\cap}$ in ${\cal H}$ while $\Pi_{\cal H_\cap}$ and $\Pi_{\cal H'}$ denote respectively the projector onto ${\cal H_{\cap}}$ and ${\cal H'}$. Then the optimal USD measurement is characterised by POVM elements of the form

\begin{eqnarray}
F^{opt}_1 & = & F^{'opt}_1 \\
F^{opt}_2 & = & F^{'opt}_2 \\
F^{opt}_? & = & F^{'opt}_? + \Pi_{\cal H_\cap}
\end{eqnarray}
\\
where the operators $F^{'opt}_1, F^{'opt}_2, F^{'opt}_?$ form a POVM $\{F_k'\}$ with support on $\cal H'$ describing the OptUSDM of a reduced problem defined by
\begin{eqnarray}
\rho'_1=\frac{1}{N_1} \Pi_{\cal H'} \rho_1 \Pi_{\cal H'}, & p'_1= \frac{N_1 p_1}{N}, & N_1 = {\rm Tr}(\rho_1 \Pi_{\cal H'}) \\
\rho'_2=\frac{1}{N_2} \Pi_{\cal H'} \rho_2 \Pi_{\cal H'}, & p'_2= \frac{N_2 p_2}{N}, & N_2 = {\rm Tr}(\rho_2 \Pi_{\cal H'}) \\
N=N_1 p_1+ N_2 p_2\; .
\end{eqnarray}

And finally, the corresponding failure probability can be written in terms of $Q'[\{F'^{opt}_k\}]$, the failure probability of the reduced problem, as
\begin{eqnarray}
Q[\{F^{opt}_k\}] &=& (1-N_1) p_1 + (1-N_2) p_2 \\
&+& NQ'[\{F'^{opt}_k\}].  \nonumber \;
\end{eqnarray}
\end{theorem}

\paragraph*{\bf Proof}
To prove the reduction theorem, we state as a first step the following lemma.

\begin{lemma}
For any positive semi-definite operators $A$ and $B$, ${\rm Tr}(AB)=0$ iff the support of the two positive semi-definite operators are orthogonal :
\begin{eqnarray}
{\rm Tr}(AB)=0 \Leftrightarrow S_A \perp S_B.
\end{eqnarray}
\end{lemma}
Indeed, if $A$ and $B$ are positive semi-definite operators, they are
diagonalisable with eigenvalues $\alpha_i >0 \quad (i=1,...,r_A)$ and $\beta_j >0 \quad (j=1,...,r_B)$. Thus
\begin{eqnarray}
{\rm Tr}(AB) &=& {\rm Tr}(\sum_i \alpha_i |\Psi_i\rangle \langle \Psi_i|  \sum_j \beta_j|\Phi_i\rangle \langle \Phi_i|) \nonumber \\
&=& \sum_{ij} \alpha_i \beta_j |\langle \Psi_i|\Phi_j \rangle|^2 \;
\end{eqnarray}
vanishes iff $\{|\Phi_i \rangle\}_i$ and $\{|\Psi_j \rangle\}_j$ span orthogonal subspaces.\\

A USD measurement described by $\{F_k\}$ satisfies ${\rm Tr}(F_1\rho_2)=0$ and ${\rm Tr}(F_2\rho_1)=0$ by definition. It means, as a consequence of Lemma 1, that $S_{F_1} \perp
{S_{\rho_2}}$ and $S_{F_2} \perp {S_{\rho_1}}$. Since ${\cal H}_{\cap}$ is a subspace of ${S_{\rho_1}}$ and ${S_{\rho_2}}$, it follows that $S_{F_1} \perp {\cal H}_{\cap}$ and $S_{F_2} \perp {\cal H}_{\cap}$. Therefore, by writing the block-matrices in $\cal{H}= \cal{H}_\cap \oplus \cal{H}'$, we have

\begin{eqnarray}
F_1 = \left( \begin{array}{cc} 0 & 0 \\ 0 & F_1' \end{array} \right)\\
F_2 = \left( \begin{array}{cc} 0 & 0 \\ 0 & F_2' \end{array} \right)\;
\end{eqnarray}
The completeness relation on $\cal H$ implies firstly
\begin{eqnarray}
F_? = \left( \begin{array}{cc} \openone_{\cal {H}_\cap} & 0 \\ 0 & F_?' \end{array} \right) = \Pi_{\cal {H}_\cap}+ F'_?
\end{eqnarray}
and secondly by the completeness relation on the reduced subspace $\cal H'$
\begin{eqnarray}
\sum_k F_k' = \openone_{\cal{H'}}.
\end{eqnarray}
It follows also that the $F_k'$ ($k=1,2,?$) are positive semi-definite and hermitian operators. Therefore, by definition,  $\{F_k'\}$ is a POVM on $\cal H'$. The fact that $F_?$ is equal to identity in the subspace $\cal H_{\cap}$ is here a direct consequence of the property of an USDM on $\cal H$. Next we will see that $\{F_k'\}$ is a POVM of a USD in $\cal H'$.\\

We define $\Pi_{\cal {H}_\cap}$ and $\Pi_{\cal{H'}}$ as the projector onto $\cal H_{\cap}$ and $\cal H'$ respectively. Thus  $\Pi_{\cal {H}_\cap} \oplus \Pi_{\cal{H'}}=\openone_{\cal H}$. For any USDM, because of the diagonal block form of the POVM, we find for $Q$
\begin{eqnarray}
Q &=& p_1 {\rm Tr}(\rho_1F_?)+p_2 {\rm Tr}(\rho_2F_?) \nonumber \\
&=& (1-N_1) p_1 + (1-N_2) p_2\\
&+& (N_1p_1+N_2p_2) ( p_1' {\rm Tr}(\rho_1'F_?')+p_2' {\rm Tr}(\rho_2'F_?')) \nonumber \\
\rm{with}\,\, \rho'_1 &=& \frac{1}{{\rm Tr}(\rho_1 \Pi_{\cal H'})} \Pi_{\cal H'} \rho_1 \Pi_{\cal H'} \\
\rho'_2 &=& \frac{1}{{\rm Tr}(\rho_2 \Pi_{\cal H'})} \Pi_{\cal H'} \rho_2 \Pi_{\cal H'}. \;
\end{eqnarray}
Here $p'_i$ ($i=1,2$) is the {\it a priori} probability corresponding to the new density matrix $\rho'_i$ ($p'_1+p'_2=1$) 
\begin{eqnarray}
p'_1= \frac{N_1 p_1}{N_1 p_1+ N_2 p_2}, \, N_1 = {\rm Tr}(\rho_1 \Pi_{\cal H'})\\
p'_2= \frac{N_2 p_2}{N_1 p_1+ N_2 p_2}, \, N_2 = {\rm Tr}(\rho_2 \Pi_{\cal H'}).\;
\end{eqnarray}

We notice that ${\cal S}_{\rho'_1} \cap {\cal S}_{\rho'_2}=0$. Moreover, ${\rm Tr}(\rho_1 F_2)=0$ implies ${\rm Tr}(\rho'_1 F'_2)=0$ and ${\rm Tr}(\rho_2 F_1)=0$ implies ${\rm Tr}(\rho'_2 F'_1)=0$. Then $\{F_k'\}$ defines a POVM describing a USDM on $\{\rho_i'\}$ in $\cal H'$. The problem is now reduced to the subspace $\cal H'$ and it remains to consider the optimality of the reduced USDM.\\

We can write $Q$ such as
\begin{eqnarray}
Q=(1-N_1) p_1 + (1-N_2) p_2 + (N_1p_1+N_2p_2) Q'\;
\end{eqnarray}
where $Q'=p_1' {\rm Tr}(\rho_1'F_?')+p_2' {\rm Tr}(\rho_2'F_?')$ is, by definition, the failure probability of discriminating unambiguously $\rho_1'$ and $\rho_2'$ in $\cal H'$.

The previous equality implies that the failure probability $Q$ is minimal iff the failure probability $Q'$ is minimal. Thus we have that $\{F_k\}$ describes an optimal USDM on $\{\rho_i\}$ $\Leftrightarrow$ $Q$ is minimal $\Leftrightarrow$ $Q'$ is minimal $\Leftrightarrow$ $\{F'_k\}$ describes an optimal USDM on $\{\rho'_i\}$. This completes the proof.

\section{Orthogonal subspaces of the supports}
\begin{figure}
\epsfbox{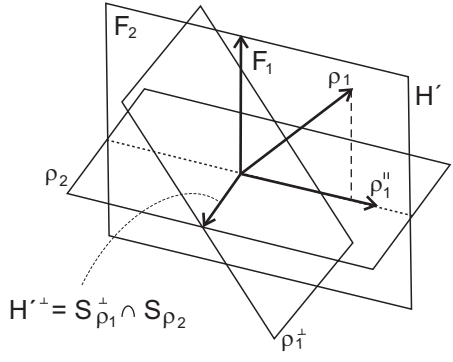}
\caption{\label{figure}\bf {Elimination of the subspace of ${\cal S}_{\rho_2}$ orthogonal to ${\cal S}_{\rho_1}$} - $\rho_1''$ denotes the orthogonal projection of $\rho_1$ onto ${\cal S}_{\rho_2}$.}
\end{figure}

We now consider the case where the supports of the two density matrices have
no common subspace. That can always be reached thanks to the previous
reduction theorem for common subspace. If there is a part of ${\cal
  S}_{\rho_2}$ orthogonal to ${\cal S}_{\rho_1}$, we can decompose ${\cal
  S}_{\rho_2}$ into this orthogonal subspace and another one. (See
Fig. \ref{figure}.) It turns out that this subspace of ${\cal S}_{\rho_2}$ orthogonal to ${\cal S}_{\rho_1}$ can be split off and leads to an unambiguous discrimination without error. The same is true for ${\cal S}_{\rho_1}$.
\begin{theorem}
Reduction Theorem for Orthogonal Subspaces\\

Suppose we are given two density matrices $\rho_1$ and $\rho_2$ in ${\cal H}$ with rank $r_1$ and $r_2$, respectively, and with their associated {\it a priori} probabilities $p_1$ and $p_2$. Assuming that their supports ${\cal S}_{\rho_1}$ and ${\cal S}_{\rho_2}$ have no common subspace, one can construct a decomposition
\begin{eqnarray}
{\cal H}={\cal H}' \oplus {\cal H}^{' \perp}
\end{eqnarray}
with ${\cal H}^{' \perp}=S_1^{\perp} \oplus S_2^{\perp}$, ${\cal S}_1^{\perp}={\cal S}_{\rho_1}^{\perp} \cap {\cal  S}_{\rho_2}$ and ${\cal S}_2^{\perp}={\cal S}_{\rho_2}^{\perp} \cap {\cal  S}_{\rho_1}$.

The solution of the optimal USDM problem can be given, with help of $\Pi_{{\cal S}^{\perp}_1}$ and $\Pi_{{\cal S}^{\perp}_2}$, the projection onto ${\cal S}_1^{\perp}$ and ${\cal S}_2^{\perp}$, respectively,  in ${\cal H}={\cal H}' \oplus {\cal H}^{' \perp}$, by
\begin{eqnarray}
F^{opt}_1 & = & F^{'opt}_1 + \Pi_{{\cal S}_2^{\perp}}\\
F^{opt}_2 & = & F^{'opt}_2 + \Pi_{{\cal S}_1^{\perp}}\\
F^{opt}_? & = & F^{'opt}_? .
\end{eqnarray}
The operators $F^{'opt}_1, F^{'opt}_2, F^{'opt}_?$ form a POVM $\{F_k'\}$ with support on $\cal H'$ describing the OptUSDM of a reduced problem defined by
\begin{eqnarray}
\rho'_1=\frac{1}{N_1} \Pi_{\cal H'} \rho_1 \Pi_{\cal H'}, & p'_1= \frac{N_1 p_1}{N}, & N_1 = {\rm Tr}(\rho_1 \Pi_{\cal H'}) \\
\rho'_2=\frac{1}{N_2} \Pi_{\cal H'} \rho_2 \Pi_{\cal H'}, & p'_2= \frac{N_2 p_2}{N}, & N_2 = {\rm Tr}(\rho_2 \Pi_{\cal H'}) \\
N=N_1 p_1+ N_2 p_2.\;
\end{eqnarray}

And finally, the corresponding failure probability can be written in terms of $Q'[\{F'^{opt}_k\}]$, the failure probability of the reduced problem as
\begin{eqnarray}
Q[\{F^{opt}_k\}] &=& NQ'[\{F'^{opt}_k\}].\;
\end{eqnarray}\\
\end{theorem}

\paragraph*{\bf Proof}
We translate the problem using a Naimark's extension and projection-valued measure (PVM). This idea is inspired by the first work of Sun {\it et al.} \cite{hillery} where an extended Hilbert space has been used. Let us repeat the Naimark theorem: Given $\{F_k\}$ as a POVM on a Hilbert space $\cal H$, it exists an embedding of $\cal H$ into a larger Hilbert space $\cal K$ such that the measure can be described by  projections onto orthogonal subspaces in $\cal K$. More precisely, there exist a Hilbert space $\cal K$, an embedding $\cal E$ such that ${\cal E }{\cal H} ={\cal K}$ and a PVM $\{E_k\}$ in $\cal K$ such that with P, the projection defined by $P{\cal K} ={\cal H}$, $F_k = PE_kP, \, \forall k$.

To the three POVM elements $F_k$ in $\cal H$ correspond three PVM elements $E_k$ in $\cal K$. The subspaces defined by $\{E_k\}$ result in a decomposition into orthogonal subspaces
\begin{eqnarray}
{\cal K} = {\cal S}_{E_1} \oplus {\cal S}_{E_2} \oplus {\cal S}_{E_?}
\end{eqnarray}
which give raise to non-orthogonal subspaces in $\cal H$ as ${\cal S}_{F_k} = P{\cal S}_{E_k}P$. We can therefore translate properties of the USD POVM to the embedding of $\cal H$ into $\cal K$.

Next we take a look at the embedding of ${\cal S}_{\rho_1}$ and ${\cal S}_{\rho_2}$ into $\cal K$ and we translate the conditions for an USDM into the embedded language. We denote embedded subspaces of $\cal K$ by the same symbol as the original subspace of $\cal H$. Then ${\rm Tr}(\rho_1 E_2)=0$ implies that ${\cal S}_{\rho_1}$ is orthogonal to ${\cal S}_{E_2}$. Similarly, we find that ${\cal S}_{\rho_2}$ is orthogonal to ${\cal S}_{E_1}$. Therefore, we can write
\begin{eqnarray}
{\cal S}_{\rho_1} \subset {\cal S}_{E_1} \oplus {\cal S}_{E_{?1}}\\
{\cal S}_{\rho_2} \subset {\cal S}_{E_2} \oplus {\cal S}_{E_{?2}}
\end{eqnarray}
where ${\cal S}_{E_{?1}}$ and ${\cal S}_{E_{?2}}$ are defined as subspaces of ${\cal S}_{E_?}$ with minimal dimension fullfiling the above decompositions in the sense that ${\cal S}_{E_{?i}}=\rm{support}(\Pi_{{\cal S}_{E_?}}{\cal S}_{\rho_i}\Pi_{{\cal S}_{E_?}})$ for $i=1,2$.

The optimality condition means in particular that no information should be obtained from the conditional states following an inconclusive result. If the two failure spaces ${\cal S}_{E_{?1}}$ and ${\cal S}_{E_{?2}}$ are different, it will be possible to distinguish the conditional states which arise from a projection onto ${\cal S}_{E_?}$ \cite{hillery}. Therefore the optimality condition implies that ${\cal S}_{E_{?1}}={\cal S}_{E_{?2}}$ and then
\begin{eqnarray}
{\cal S}_{E_{?}}={\cal S}_{E_{?1}}={\cal S}_{E_{?2}}.
\end{eqnarray}
In the framework of the Naimark extension, this condition translates as follows : the equality of ${\cal S}_{E_{?1}}$ and ${\cal S}_{E_{?2}}$ implies that a subspace ${\cal S}_1^{\perp}={\cal S}_{\rho_1}^{\perp} \cap {\cal  S}_{\rho_2}$ satisfies ${\cal S}_1^{\perp} \subset {\cal S}_{E_2}$ in order to assure that the overlap between any state in ${\cal S}_1^{\perp}$ and any state in ${\cal S}_{\rho_1}$ will be zero. Similarly, ${\cal S}_2^{\perp} \subset {\cal S}_{E_1}$.


Then it exists a subspace ${\cal H}_2$ in ${\cal S}_{E_2}$ such that we can write ${\cal S}_{E_2}={\cal S}_1^{\perp} \oplus {\cal H}_2$. In the same way, ${\cal S}_{E_1}={\cal S}_2^{\perp} \oplus {\cal H}_1$ with ${\cal H}_1$ in ${\cal S}_{E_1}$. It follows that
\begin{eqnarray}
{\cal S}_{\rho_1} \subset {\cal S}_2^{\perp} \oplus {\cal H}_1 \oplus {\cal S}_{E_{?1}}\\
{\cal S}_{\rho_2} \subset {\cal S}_1^{\perp} \oplus {\cal H}_2 \oplus {\cal S}_{E_{?2}}.
\end{eqnarray}
The fact that ${\cal S}_2^{\perp} \subset {\cal S}_{\rho_1}$ implies that
\begin{eqnarray}
{\cal S}_{\rho_1} = {\cal S}_2^{\perp} \oplus {\cal H}'_1,
\end{eqnarray}
with ${\cal H}'_1 \subset {\cal H}_1 \oplus {\cal S}_{E_{?1}}$. In the same way, with ${\cal H}'_2 \subset {\cal H}_2 \oplus {\cal S}_{E_{?2}}$,
\begin{eqnarray}
{\cal S}_{\rho_2} = {\cal S}_1^{\perp} \oplus {\cal H}'_2. 
\end{eqnarray}


The orthogonal projection $E_1$ then can be decomposed into a sum of orthogonal projectors as $\Pi_{{\cal S}^{\perp}_2} + \Pi_{{\cal H}'_1}$ and the orthogonal projection $E_2$ as $\Pi_{{\cal S}^{\perp}_1} + \Pi_{{\cal H}'_2}$. These projectors are mapped into $\cal H$ via the projection $P$ as  $P\Pi_{{\cal S}_i^{\perp}}P = \Pi_{{\cal S}_i^{\perp}}$. We define $F'_i=P\Pi_{{\cal H}_i}P \quad \forall i=1,2$ so that
\begin{eqnarray}
F_1 & = & F'_1 + \Pi_{{\cal S}_2^{\perp}}\\
F_2 & = & F'_2 + \Pi_{{\cal S}_1^{\perp}}.\;
\end{eqnarray}
with ${\cal S}_{F'_1} \subset {({\cal S}_1^{\perp})}^{\perp}$ and ${\cal S}_{F'_2} \subset {({\cal S}_2^{\perp})}^{\perp}$. Moreover, ${\cal S}_{F_1} \perp {\cal S}_{\rho_2}$ then ${\cal S}_{F_1} \perp {\cal S}_1^{\perp}$ and, similarly, ${\cal S}_{F_2} \perp {\cal S}_2^{\perp}$. Then $F'_1$ and $F'_2$ have support on a subspace ${\cal H}'$, which is the complementary orthogonal subspace of ${\cal H}^{'\perp}={\cal S}_1^{\perp} \oplus {\cal S}_2^{\perp}$.

Therefore in ${\cal H}={\cal H}' \oplus {\cal S}_1^{\perp} \oplus {\cal S}_2^{\perp}={\cal H}'\oplus {\cal H}^{' \perp}$, we find
\begin{eqnarray}
F_1 = \left( \begin{array}{ccc} F'_1 & 0 & 0 \\ 0 & \openone_{{\cal S}_1^{\perp}} & 0 \\ 0 & 0 & 0 \end{array} \right)\\
F_2 = \left( \begin{array}{ccc} F'_2 & 0 & 0 \\ 0 & 0 & 0 \\ 0 & 0 & \openone_{{\cal S}_2^{\perp}} \end{array} \right).\;
\end{eqnarray}
From here, we will follow the same argumentation as we used in the proof of theorem 1. The completeness relation on $\cal H$ implies firstly
\begin{eqnarray}
F_? = \left( \begin{array}{ccc} F'_? & 0 &0 \\ 0& 0 & 0 \\0 & 0 & 0 \end{array} \right)
\end{eqnarray}
and secondly the completeness relation on the reduced subspace $\cal H'$
\begin{eqnarray}
\sum_k F_k' = \openone_{\cal{H'}}.
\end{eqnarray}
It follows also that the $F_k'$ ($k=1,2,?$) are positive semi-definite operators. Therefore, by definition, $\{F_k'\}$ is a POVM on $\cal H'$. 

For any USDM, because of the diagonal block form of the POVM, we find for $Q$
\begin{eqnarray}
Q &=& p_1 {\rm Tr}(\rho_1F_?)+p_2 {\rm Tr}(\rho_2F_?) \\
&=& (N_1p_1+N_2p_2) ( p_1' {\rm Tr}(\rho_1'F'_?)+p_2' {\rm Tr}(\rho_2'F_?')) \nonumber \\
\rm{with}\,\, \rho'_1 &=& \frac{1}{{\rm Tr}(\rho_1 \Pi_{\cal H'})} \Pi_{\cal H'} \rho_1 \Pi_{\cal H'} \\
\rho'_2 &=& \frac{1}{{\rm Tr}(\rho_2 \Pi_{\cal H'})} \Pi_{\cal H'} \rho_2 \Pi_{\cal H'}. \;
\end{eqnarray}
Here $p'_i$ ($i=1,2$) is the {\it a priori} probability corresponding to the new density matrix $\rho'_i$ ($p'_1+p'_2=1$) 
\begin{eqnarray}
p'_1= \frac{N_1 p_1}{N_1 p_1+ N_2 p_2}, \, N_1 = {\rm Tr}(\rho_1 \Pi_{\cal H'})\\
p'_2= \frac{N_2 p_2}{N_1 p_1+ N_2 p_2}, \, N_2 = {\rm Tr}(\rho_2 \Pi_{\cal H'}).\;
\end{eqnarray}
Moreover, ${\rm Tr}(\rho_1 F_2)=0$ implies ${\rm Tr}(\rho'_1 F'_2)=0$ and ${\rm Tr}(\rho_2 F_1)=0$ implies ${\rm Tr}(\rho'_2 F'_1)=0$. Then $\{F_k'\}$ defines a POVM describing a USDM on $\{\rho_i'\}$ in $\cal H'$.\\

We can rewrite the failure probability $Q$ as
\begin{eqnarray}
Q=(N_1p_1+N_2p_2) Q'\;
\end{eqnarray}
where $Q'=p_1' {\rm Tr}(\rho_1'F_?')+p_2' {\rm Tr}(\rho_2'F_?')$ is, by definition, the failure probability of discriminating unambiguously $\rho_1'$ and $\rho_2'$ in $\cal H'$ with {\it a priori} probabilities $p'_1$ and $p'_2$, respectively.

And again, we have that $\{F_k\}$ describes an optimal USDM on $\{\rho_i\}$ $\Leftrightarrow$ $Q$ is minimal $\Leftrightarrow$ $Q'$ is minimal $\Leftrightarrow$ $\{F'_k\}$ describes an optimal USDM on $\{\rho'_i\}$. This completes the proof.

\section{Consequences and generalization}
At this point, it is useful to introduce a notation to summarise our knowledge about the USD of two density matrices. We have ${\cal H}= {\cal S}_{\rho_1} + {\cal S}_{\rho_2}$ then ${\rm dim}\left({\cal H}\right) = {\rm dim}\left({\cal S}_{\rho_1}\right) + {\rm dim}\left({\cal S}_{\rho_2}\right) - {\rm dim}\left({\cal S}_{\rho_1} \cap {\cal S}_{\rho_2}\right)$. It implies, by denoting the dimension of the Hilbert space $\cal H$ as $d$, that the respective ranks of the density matrices satisfies
\begin{eqnarray}
r_1 + r_2 \ge d.
\end{eqnarray}
For example, the case of two density matrices of the same rank $(n-1)$ in an Hilbert space of dimension $n$ described by Rudolph {\it et al.} \cite{rudolph} can be written as ``$\left(n-1\right)+\left(n-1\right) > n$'' while the USD between one pure state and a mixed state described by Bergou {\it et al.} \cite{bergou} can be characterised as the ``$1+n = (n+1)$'' case.\\

Now we discuss interesting consequences to the two above theorems. First of all, the first theorem corresponds to the elimination of the common subspace. A common subspace is present when $r_1+r_2 >d$ holds. Its dimension is $d_{\cap}=r_1+r_2-d$. Therefore, after elimination of that subspace, we end up in the case $r'_1+r'_2=d'$ with $r'_1=r_1-d_{\cap}$ and similarly for $r'_2$ and $d'$. Then, we can reduce the Rudolph's case of discriminating unambiguously two density matrices of the same rank $(n-1)$ in an Hilbert space of dimension $n$ to the ``$1+1=2$'' case of two pure states because the common subspace is ($n-2$)-dimensional. Rudolph {\it et al.} \cite{rudolph} already noticed it in their paper. The reduction is constructive given $\rho_1$ and $\rho_2$.

The second theorem corresponds to the elimination of the orthogonal part of one support with respect to the other, i.e., ${\cal S}_{\rho_1}^{\perp} \cap {\cal  S}_{\rho_2}$ and ${\cal S}_{\rho_2}^{\perp} \cap {\cal  S}_{\rho_1}$. The non-empty subspaces ${\cal S}_{\rho_1}^{\perp} \cap {\cal S}_{\rho_2}$  and ${\cal S}_{\rho_2}^{\perp} \cap {\cal S}_{\rho_1}$ can be found systematically. For example, ${\cal S}_{\rho_1}^{\perp} \cap {\cal S}_{\rho_2}$ can be found by projecting ${\cal S}_{\rho_1}$ onto ${\cal S}_{\rho_2}$ and then by taking the complementary orthogonal subspace in ${\cal S}_{\rho_2}$ of that projection. As a matter of fact, this assures that we can reduce the general USD problem always to that of two density matrices of the same rank $r$, $r \le \min(r_1, r_2)$, in a Hilbert space of $2r$ dimensions. Indeed, if after the reduction the rank of $\rho_2'$ is bigger than the rank of $\rho_1'$, then the subspace ${\cal S}_{\rho'_1}^{\perp} \cap {\cal S}_{\rho'_2}$ is at least of dimension $r'_2-r'_1$ and can be eliminated.
With the help of the two above theorems, we can reduce any problem of discriminating unambiguously two density matrices $\rho_1$ and $\rho_2$, with rank $r_1$ and $r_2$ respectively, in a Hilbert space $\cal H$, into a problem of discriminating unambiguously two density matrices $\rho_1'$ and $\rho_2'$ with rank $r$ ($r \le \min (r_1, r_2)$) in $\cal H' \subset H$, a 2$r$-dimensional Hilbert space. The reduction is constructive. The first theorem allows us to split off the common subspace and the second theorem leads to the reduce problem of discriminating unambiguously two density matrices of the same rank.

As a consequence, we can reduce, for example, the problem of USD between a pure state and a density matrix, the ``$1+n=(n+1)$'' case, to the problem of discriminating unambiguously two pure states, that is to say the ``$1+1=2$'' case, by splitting off ${\cal S}_{\rho_1}^{\perp} \cap {\cal S}_{\rho_2}$ of dimension $(n-1)$. The two states are the original pure state $\rho_1$ and the unit vector corresponding to the projection of the original pure state onto the support of the mixed state $\rho_2$.

It implies that the only two exact solutions of OptUSD between mixed states that are known so far, on one hand, Bergou {\it et al.} \cite{bergou} and, on the other hand, Rudolph {\it et al.} \cite{rudolph}, can be derived from the ``$1+1=2$'' case of Jaeger and Shimony\cite{jaeger}.\\

It is also interesting to note that the dimension of the failure space can not be greater than the lowest rank of the involved density matrices. First we have $F_? = PE_?P$ so that ${\rm dim}({\cal S}_{F_?}) \le {\rm dim}({\cal S}_{E_?})$. Second the dimension of ${\cal S}_{E_{?i}}$ can not be greater than $r_i$ because ${\cal S}_{E_{?i}}=\rm{support}(\Pi_{{\cal S}_{E_?}}{\cal S}_{\rho_i}\Pi_{{\cal S}_{E_?}})$, for $i=1,2$, and ${\cal S}_{E_{?}}={\cal S}_{E_{?1}}={\cal S}_{E_{?2}}$. Therefore
\begin{eqnarray}
{\rm dim}{\cal S}_{E_?} \le \min_i {\rm dim}{\cal S}_{\rho_i}.
\end{eqnarray}
This result looks natural considering that we can finally reduce any problem of discriminating two density matrices with rank $r_1$ and $r_2$, respectively, to the problem of discriminating two density matrices of the same rank $r$, $r \le \min_i r_i$.\\

Finally, a generalization to more than two density matrices can be achieved. Considering $N$ density matrices $\rho_k \,  (k=1...N)$ with {\it a priori} probabilities $p_k$, we can construct $N$ pairs of density matrices ${\tilde \rho_1}=\rho_i$, $i \in [1,..,N ]$ and ${\tilde \rho_2}=\frac{\sum_{j=1, j \ne i}^N p_j\rho_j}{1-p_i}$, with ${\tilde p}_1=p_i$, ${\tilde p}_2=1-p_i$, and apply the two reduction theorems to these two density matrices in the following sense. We notice that ${\tilde \rho_2}$ has no physical meaning. Actually, as soon as a common subspace between any ${\cal S}_{\tilde \rho_1}$ and ${\cal S}_{\tilde \rho_2}$ exists, we can split it off from all the ${\cal S}\rho_i$'s because if we cannot discriminate unambiguously this part of the support of ${\tilde \rho_1}$ and ${\tilde \rho_2}$ then we can not discriminate unambiguously between this part of the support of all the $\rho_j$. The second theorem must be used more carefully. As soon as a subspace of ${\cal S}_{\tilde \rho_1}$ is orthogonal to ${\cal S}_{\tilde \rho_2}$ (${\cal S}_{\tilde \rho_2}^{\perp} \cap {\cal  S}_{\tilde\rho_1} \ne \{0\}$), we can eliminate it from the problem because it is orthogonal to the supports of all the $\rho_j$, $j \ne i$. However we cannot eliminate a subspace of ${\cal S}_{\tilde \rho_2}$ orthogonal to ${\cal S}_{\tilde \rho_1}$ (${\cal S}_{\tilde \rho_1}^{\perp} \cap {\cal  S}_{\tilde\rho_2} \ne \{0\}$) because we know nothing about the orthogonality of this subspace for all the states in $\tilde \rho_2$. In other words, we can only reduce the density matrix $\rho_i$ corresponding to ${\tilde \rho_1}$. 

\section{Conclusion}
We have shown that the problem of discriminating unambiguously any two density matrices can be reduced to the problem of discriminating unambiguously two density matrices of the same rank $r$ in a Hilbert space of 2$r$ dimensions. Firstly, we can split off any common subspace of the supports and, secondly, we can eliminate the part of the support of $\rho _2$ which is orthogonal to the support of $\rho _1$ and {\it {vice versa}}.

Finally, all the previous exact solutions of USD between two mixed states can be reduced to the the ``$1+1=2$'' case that is to say, the unambiguous discrimination of two pure states.

To conclude, given any two density matrices, we can explicitly reduce the problem ``$r_1 + r_2 \ge d$'' to a problem ``$r+r=2r$'', where $r \le r_1 \le r_2$, which is the discrimination between two density matrices sharing the same rank. For the theoretical side, it implies that the only relevant cases to study are the ``$r+r=2r$'' cases, $\forall r $. The next step should be to solve analytically the general ``$2+2=4$'' case.

\begin{center}
{\bf Acknowledgements}
\end{center}

We thank Aska Dolinska, Marcos Curty and Peter van Loock for very useful discussions. This work was supported by the DFG under the Emmy-Noether programme, the EU FET network RAMBOQ (IST-2002-6.2.1) and the network of competence QIP of the state of Bavaria (A8).


\begin{thebibliography}{99}
\bibitem{dieks} D. Dieks, Phys. Lett. A {\bf 126}, 303 (1988)
\bibitem{ivanovic} I.D. Ivanovic, Phys. Lett. A, {\bf 123}, 257 (1987)
\bibitem{peres} A. Peres, Phys. Lett. A, {\bf 128}, 19 (1988)
\bibitem{jaeger} G. Jaeger and A. Shimony, Phys. Lett. A, {\bf 197}, 83 (1995)
\bibitem{terno} A. Peres and Terno, J. Phys. A, {\bf 31}, 7105 (1998)
\bibitem{chefles} A. Chefles and S.M. Barnett, Phys. Lett. A, {\bf 250}, 223 (1998)
\bibitem{hillery} Y. Sun, J. A. Bergou and M. Hillery, Phys. Rev. A {\bf 66}, 032315 (2002) 
\bibitem{bergou} J.A. Bergou, U. Herzog and M. Hillery, arXiv:quant-ph/0209007 (2002)
\bibitem{rudolph} T. Rudolph, R. W. Spekkens and P. S. Turner, arXiv:quant-ph/0303071 (2003)
\bibitem{fiurasek} J. Fiurasek and M. Jezek, Phys. Rev. A, {\bf 67}, 012321 (2003)
\bibitem{eldar} Y.C. Eldar, Phys. Rev. A {\bf 67}, 042309 (2003)
\bibitem{helstrom} C.W. Helstrom, ``Quantum detection and estimation theory'', Academic Press, New York (1976)
\bibitem{kraus} K. Kraus, ``States, Effects, and Operations'', Springer, Berlin (1983)
\end{thebibliography}
\end{document}